\documentclass[10pt,twocolumn]{article}

\usepackage{amssymb}
\usepackage{amsmath}
\usepackage[english]{babel}
\usepackage{bm}
\usepackage{fontenc}
\usepackage[a4paper,left=1.825cm,right=1.825cm,top=1.825cm,bottom=1.825cm]{geometry}
\usepackage{graphicx}
\usepackage{hyperref}
\usepackage[utf8]{inputenc}

\begin{document}

\twocolumn[
\begin{center}
\Large{\textbf{Approximate Acoustic Boundary Conditions \\ in the Time-Domain using Volume Penalization}}\bigskip

\normalsize{Mathias Lemke and Julius Reiss}\smallskip

Fachgebiet Numerische Fluiddynamik, Technische Universität Berlin \\ Müller-Breslau-Strasse~15, 10623 Berlin, Germany 
\end{center}

\begin{center}
    \begin{minipage}{0.89\textwidth}
        \begin{footnotesize}
            Immersed boundary methods allow describing complex objects on simple Cartesian grids in time-domain simulations.
            The penalization technique employed here is a physically motivated Brinkman method that models objects as porous material by including a friction term and an effective volume.
            We investigate how the approach can mimic different acoustic boundary conditions. 
            It is validated concerning different acoustic setups, including rigid walls and various absorber configurations.
        \end{footnotesize}
    \end{minipage}
\end{center}
\bigskip

]

\section{\label{sec_1} Introduction}

Acoustic time-domain simulations allow describing effects, which are difficult to include in geometrical acoustics, like diffraction or non-constant sound propagation velocities.
These advantages often justify the high numerical effort compared with classical methods. 
That holds especially since the increasing computational resources permit practical calculations for frequencies above the Schroeder frequency on standard workstations.

Time-domain simulations can be based on the wave equation \cite{KowalczykWalstijn2011}, the non-linear Euler equations \cite{SteinStraubeSesterhennWeinzierlLemke2019,SteinStraubeWeinzierlLemke2020} or the acoustic equations as its linearized form.  
A variety of numerical methods has been utilized, like the finite-difference (FD) \cite{Botteldooren1994}, the finite element (FEM) \cite{Craggs1994}, the finite volume (FVM) \cite{Bilbao2013} or the discontinuous Galerkin method (DG) \cite{PindJeongEngsigKarupEtAl2020}.

In all these approaches, the boundary conditions are in particular decisive for the quality of the simulations \cite{Vorlaender2013}. 
Acoustic boundary conditions are often characterized by their impedance. 
However, the complex-valued variable is primarily suitable for frequency-based analyzes. 
A transfer to the time domain is difficult and the subject of current research \cite{FungJu2004,LiLiTam2011,PindEngsigKarupJeongEtAl2019,PindJeongEngsigKarupEtAl2020}. 

This article presents an immersed boundary method in the time domain, implemented via finite-differences, that is physically motivated and able to model objects with typical acoustic impedances.
The approach is easy to use, computationally efficient, fully parallelizable, and does not need particular boundary adaptations of the grid.

Immersed boundary methods replace the enforcement of boundary conditions on grid lines or element boundaries by additional force-like terms in the governing equations. 
Various methods exist:
The Brinkman volume penalization models objects as porous material and approximates solid objects for a vanishing porosity \cite{LiuVasilyev2007}. 
Among others, the approach is used in aeroacoustics \cite{KomatsuIwakamiHattori2016}.
Often two parameters model the effect of the porous material, a linear friction relative to the material (Darcy term) for the fluid velocity (becoming the particle velocity in a pure acoustic case), and the effective volume $\phi$, commonly called porosity in this context. 
The effective volume is included in the governing (flow) equations for flows differently by different authors as detailed in \cite{KevlahanDubosAechtner2015,Reiss2021}. 
The equations of \cite{Reiss2021} agree with a model derived from a two phase flow description \cite{KemmGaburroTheinDumbser2020} and similar for mechanical waves in earth crust with complex typologies by 
\cite{TavelliDumbserCharrierRannabauerWeinzierlBader2019}.
The acoustic reflectivity of porous material is investigated in the time domain by \cite{WilsonOstashevCollierSymonsAldridgeMarlin2007}.  
Consistent treatment of the effective volume allows varying these two parameters largely independently over a wide range and thereby mimicking different boundary conditions \cite{Reiss2021}.
Here an extended investigation of this approach for various acoustic boundary conditions is presented.  

The manuscript is structured as follows:
Section \ref{sec_2} introduces the governing equations.
In Sec.~\ref{sec_3} their numerical discretization is discussed.
Section \ref{sec_4} presents validation results for rigid wall and absorber configurations as well as reactive boundaries in the form of Helmholtz resonators for different setups from 1-D to 3-D.
A summary of the findings is given in Sec.~\ref{sec_summary}.
In the appendix a discussion on the use of the volume penalization approach within the wave equation is provided.

\section{\label{sec_2} Governing Equations}

The Euler-equations \cite{LandauLifshitz1987} are used as governing equations for acoustics as they comprise sound formation and propagation, including non-linear effects.
The Brinkman penalization is introduced into the Euler-equations by an effective volume $\bm{\phi}$ \cite{Reiss2021} and a Darcy term proportional to $\chi$ \cite{KomatsuIwakamiHattori2016} on the right-hand-side to enable suitable impedance boundary conditions.
The terms are highlighted in bold font.
\begin{align}
	\bm{\phi} \partial_t ( \rho) + \partial_{x_i}( \bm{\phi} \rho u_i)                                &= 0  \label{eqn_mass} \\ 
	\bm{\phi} \partial_t ( \rho u_j) + \partial_{x_i}( \bm{\phi} \rho u_i u_j) + \bm{\phi} \partial_{x_j}  p  &=  \bm{\phi \chi (u_j^t - u_j)}\label{eqn_momentum}  \\  
 	\bm{\phi} \partial_t ( \rho e_t )  + \partial_{x_i}( \bm{\phi} \varrho u_i e_t + \bm{\phi}u_i p) &= 0  \label{eqn_energy}
\end{align}    
Therein, $\rho$ denotes the density, $u_j$ the velocity in $x_j$-direction, $e_t$ the specific total energy, $p$ the pressure and $\gamma$ the heat capacity ratio. 
The value $u_j^t$ corresponds to the velocity of the modeled boundary/object and is zero in the further course of this manuscript.
We do not explicitly emphasize that the variables are functions of space and time for the sake of brevity.
The sum convention applies for $i,j = [1,2,3]$. 
Assuming a constant heat capacity the energy equation can be reformulated using $e_t = (p/\rho) \cdot 1/(\gamma - 1) + (u_ju_j)/2$ \cite{LemkeReissSesterhenn2014} resulting in
\begin{align}
    \bm{\phi} \partial_t(p) + \gamma \partial_{x_i}(\bm{\phi} u_i p) 
    - (\gamma -1 )  \bm{\phi} u_i  \partial_{x_i} p
    &=  0. \label{eqn_pressure}
\end{align}    
The acoustic equations are provided in the appendix. 

\paragraph{Effective volume $(\bm \phi)$}

The inclusion of the effective volume $\bf \phi$ can be interpreted as an volume fraction $\bm{\phi}(x_i) = V_\mathrm{fluid}/V_\mathrm{total}$ caused by the presence of a porous medium.
Thus, the value of $\bm \phi$ varies between $0$ and $1$.
With $\bm \phi = 0$ the Euler equations \eqref{eqn_mass}-\eqref{eqn_energy} degenerate.
This situation is avoided by choosing a small but finite value for $\phi$ in the simulations.
With $\bm \phi = 1$ the unmodified Euler equations are recovered, a spatial constant $\phi$ can be factored out.
Negative values or values greater than one are non-physical and are excluded.
A time-depended $\phi$ is discussed in \cite{KemmGaburroTheinDumbser2020,Reiss2021} and not considered here.

Another interpretation of the effective volume $\bm \phi$ is a reduced cross-section of a stream tube.
We will exploit this second interpretation when modelling Helmholtz resonators in Sec.~\ref{ssec_1d_helmholtz}.

An extensive discussion of the terms is carried out in \cite{Reiss2021,KemmGaburroTheinDumbser2020}.
In particular, it is argued that the presence of the effective volume does not change the local speed of sound and interferes only slightly with the eigenvalues of the governing equations.
Thus, existing aeroacoustic simulation programs can be reused by modifying the governing equations.

\paragraph{Darcy term $(\bm{\phi \chi (u_j^t - u_j)})$}

The momentum equations are expanded by a penalization of form $(\bm{\phi \chi (u_j^t - u_j)})$
with $\bm{\chi}(x_i) = \chi_{x}(x_i) \cdot \chi_s(x_i)$ controlling its spatial location $\chi_{x}$ and strength $\chi_s$.
This is referred to as Darcy-term in the following.   
The expression can be interpreted as velocity damping caused by the presence of a porous material.
The target velocity $u_j^t$ corresponds to the velocity of the material and is chosen to zero in the following.
While the location parameter $\chi_x$ varies between $0$ and $1$, the strength parameter $\chi_s$ varies between $0$ and $\infty$.
In the limit $\chi_s \rightarrow \infty$ the Euler equations degenerate.
For $\chi = 0$ the force term vanishes. 
Larger values of $\chi$ result in larger negative Eigenvalues of the right-hand-side operator, which has a corresponding influence on the stability of the numerical simulation or requires substantially smaller time steps for explicit time marching.

Regardless of this, it is possible to model solid/fully reflective or semi-permeable boundaries with sufficiently large values of $\chi$ \cite{KomatsuIwakamiHattori2016,LemkeCitroGiannetti2021}, analogous to the effective volume.
However, in the following, $\phi$ is used to model solid walls and not a large Darcy term. 
In the validation examples, the values of $\chi$ used for modeling practically relevant acoustic boundaries do not lead to restrictions in terms of stability and the time step, respectively.

The term usually added to the energy equation is dropped, as in the short computational time, no significant heat effects are expected \cite{LiuVasilyev2007,LemkeCitroGiannetti2021}.

\section{\label{sec_3} Numerical Discretization}

\subsection{Space discretization}

The governing partial differential equations (\ref{eqn_mass}-\ref{eqn_energy}) are discretized using the finite difference approach in the time-domain (FDTD) on an equidistant Cartesian grid.
The spatial derivatives are approximated by finite differences.
Explicit schemes, schemes that are optimized concerning the transmission behavior \cite{TamWebb1993}, implicit schemes \cite{Lele1992} that include the solution of a linear system of equations, or spectral methods can be employed.
In principle, all finite difference (FD) schemes are suitable for discretizing the governing equations.

The variables $\phi(x_i)$ and $\chi(x_i)$ are chosen to define an acoustic (and fluid dynamic) boundary condition. 
Both are field variables and can thereby have different values for each grid point. 
The values are motivated by physics and can be estimated from material properties, see below.
The effective volume in $\phi$ is identified as the volume not occupied by a porous material or a resonator.
The variable $\chi$ corresponds to the flow resistivity. 
Initial guesses by the physical quantities provide approximations that are often good enough for basic applications.
Thus, the Brinkman penalization is not a first-principles approach but an approximate description.

\subsection{Time discretization}

Throughout this manuscript, the time is discretized by the standard explicit Runge-Kutta-4 (RK4)  method \cite{SchwarzKoeckler2009}. 
As an explicit method, it is easy to implement but restricts the maximal time-step.
The eigenvalues of the spatially discretized equations scaled by the time-step must be within the so-called stability region \cite{SchwarzKoeckler2009} of the method.  
For the inviscid problems of acoustics, the time-step can be estimated by the CFL number $\frac{c \Delta t}{\Delta x}\leq \mathrm{CFL_{max}} $, where $\Delta x,\,\Delta t $  are the spatial and temporal discretization, $c$ is the speed of sound and $\mathrm{CFL_{max}}$ is a limit depending on details of the discretization which is usually of the order of one.

Large values of $\chi_s$ increase the eigenvalues and thus reduce the permitted time step $\Delta t$. 
Sharp spatial changes in $\phi$  do have a similar effect, whereas a smoothing to distribute the change over a few grid points avoids such a restriction \cite{Reiss2021}. 

Larger values of the Darcy term or sharper changes of $\phi$ could be handled by an implicit or semi-implicit method as used in \cite{BoironChiavassaDonat2009}. 
A large $\chi_s$ corresponds to simply setting the target values inside rigid objects. 
This reduces the smoothness of the solution having a detrimental effect on the discrete solution \cite{EngelsKolomenskiySchneiderSesterhenn2015,vanyenKolomenskiySchneider2014}. 
For realistic porous materials, the values of $\chi$ are low, and no restriction of the time-step is expected. 

To ensure numerical stability, a spatial filter can be used.
In some examples in Sec.~\ref{sec_4}, a compact filter following \cite{GaitondeVisbal2000} is employed. 
However, a standard filter can result in a conservation defect for a non-constant effective volume $\phi$, which is usually not essential for acoustic simulations. 
A conservative filter with varying $\phi$ is discussed in \cite{Reiss2021}.

\subsection{Blending functions}

The values $\phi$  need to be smoothed to avoid numerical problems like stiffness \cite{Reiss2021}. 
In this publication the smoothing is done by a hyperbolic tangent  function, e.g. for a wall 
at position $x_0$ the effective volume is given by 
\begin{align}
 \phi(x) = 1 - (1-\phi_\epsilon) (\tanh( (x-x_0)/\delta) + 1 )/2 .   
\end{align}
Here $\delta$ determines the smoothing width and $\phi_\epsilon $ the residual volume of the wall. 
The smoothing influences the acoustic behavior of the boundary. 
It is tested with an adiabatic Gaussian pulse $p= \exp( -x^2/\sigma_\mathrm{pulse}^2) $, with $\sigma_\mathrm{pulse} = 8\Delta x  $, with the grid spacing $\Delta x $. 
The results are computed using a standard fourth-order FD derivative and the  RK4  with a CFL number of $\approx 0.7$.

\begin{figure}
    \centering
    \includegraphics[width = \linewidth]{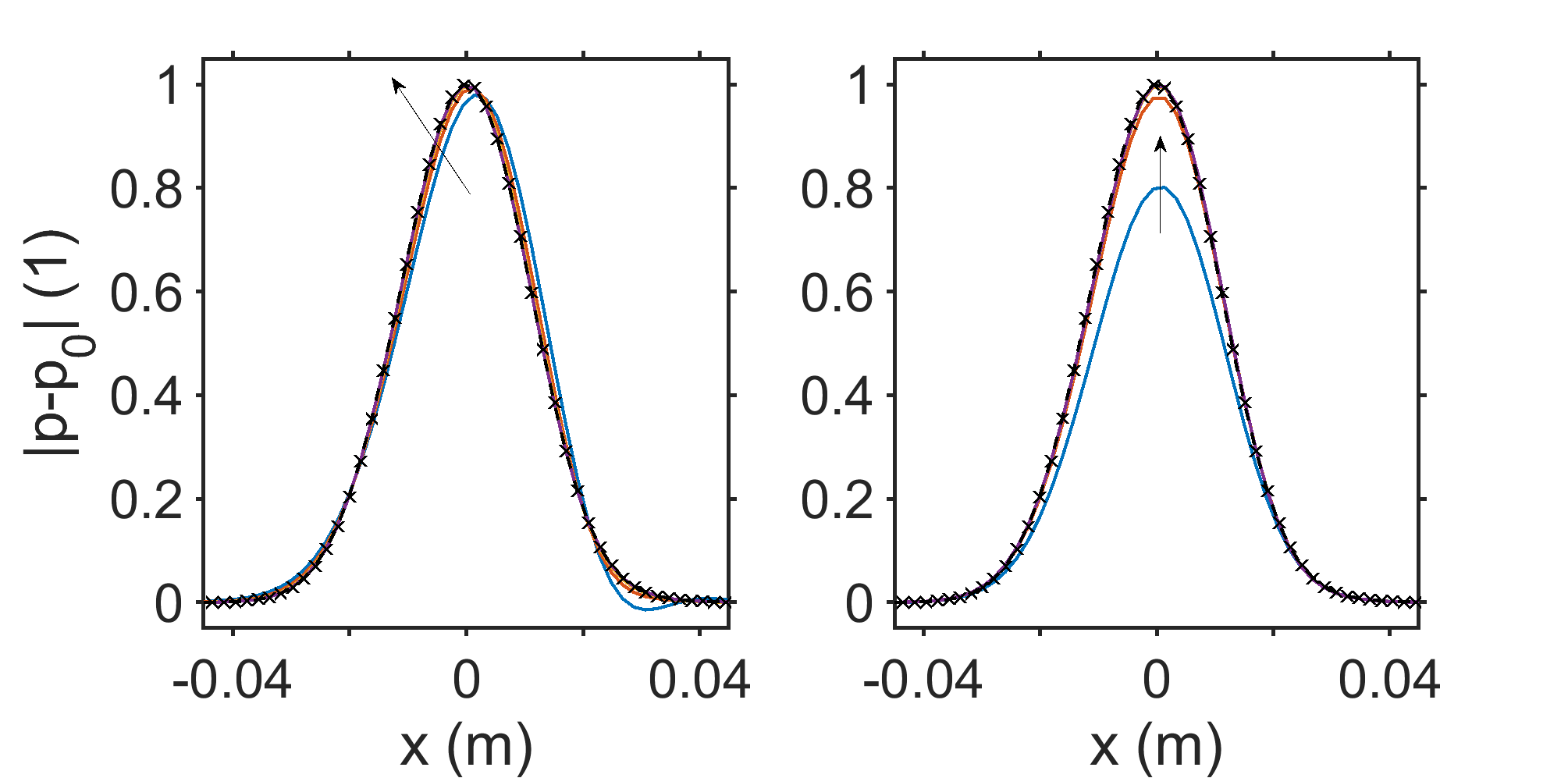}
    \caption{(color online) Reflected pressure pulse in comparison to a reference (black, dashed with crosses representing the grid points) created by a mirror pulse, both normalized with respect to the ambient pressure $p_0$ and the maximum amplitude. 
    (Left) Change of the reflected pulse with $\delta= 2.0\Delta x,\,1.5\Delta x,\,1.0\Delta x,\,0.5\Delta x,\,$ using $\phi_\epsilon= 10^{-3}$. The phase error increases with $\delta$.
    (Right) Change of the reflected pulse for $\phi_\epsilon = 10^{-1},\,10^{-2},\,10^{-3},\,10^{-4}$ using $\delta = 1 \Delta x$. For an increasing $\phi_\epsilon$ the reflected wave becomes smaller in amplitude.
    \label{fig_blending}}
\end{figure}

The smoothing of $\phi$ yields a different penetration depth for different wavenumbers, becoming larger with larger $\delta$,  visible as a dissipation-like error, see Fig.~\ref{fig_blending}, left. 
It can be shown that the error is dominantly a phase error \cite{Reiss2021} for small $\phi_\epsilon$. 
A small value of $\phi_\epsilon$ is desired for a rigid wall.
If it is not close to zero only a partial reflection of the acoustic wave results, see Fig.~\ref{fig_blending}, right.    
In combination with a suitable Darcy term, this can be used to create partially reflecting walls.

A combination of a small $\phi_\epsilon$ and a small $\delta$, is implied for a rigid wall.
However, such a combination makes the equations stiff and thereby severely restricts the time step. 
If a total reflection is not essential (for example, to reproduce theoretical cases),  a value of $\phi_\epsilon \sim 10^{-3}$ and a $\delta = 0.5 \Delta x$ seem a good compromise, as this allows to keep nearly the original time step.
For realistic rigid walls, even a lower reflectivity and thereby a larger $\phi_\epsilon$ seems adequate.    

Functions other than the hyperbolic tangent function can be used for smoothing. 
The error function $\mathrm{erf}$ was tested, but found to produce stiffer problems. 
No extensive study was carried out on this aspect since the hyperbolic tangent produces satisfactory results. 

\section{\label{sec_4} Numerical validation examples}

\begin{figure}
    \centering
    \includegraphics[width = 1\linewidth]{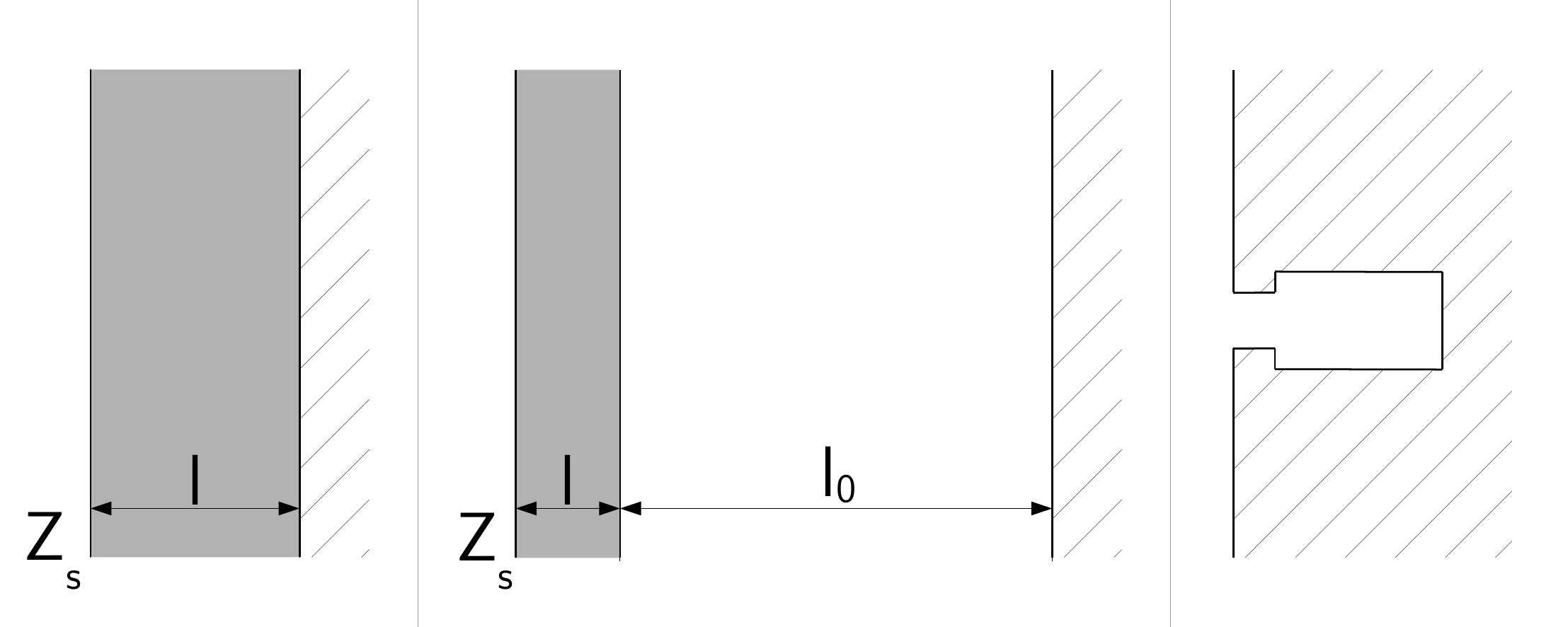}
    \caption{Acoustic configurations: (A1, left) Porous material, mounted on a rigid wall. (A2, center) Porous material mounted in front of an air cavity. (B, right) Model representation of a resonator.}
    \label{fig_a1_a2_setup}
\end{figure}

In the following, the Brinkman penalization approach for the representation of acoustic wall boundary conditions is validated using different examples. 
This includes various configurations with absorbers and rigid walls, examined in several dimensions. 
Figure \ref{fig_a1_a2_setup} shows the basic structure of the absorbers and the resonator examined below.
The key figures of the absorbers are shown in Tbl.~\ref{tab_a1_a2_setup}. 

\begin{table}[tb]
\caption{
\label{tab_a1_a2_setup}
Parameters of the absorber configuration used for validation.\bigskip}
\centering
\begin{tabular}{cccc}
\hline
\hline
~ & \tiny{material thickness} & \tiny{air cavity depth} & \tiny{flow resistivity} \\
~ & $l$ (m) & $l_0$ (m) & $\sigma$ \\
\hline
A1 & 0.1    & 0     & 3000 \\
A2 & 0.05   & 0.15  & 14400 \\
\hline
\hline
\end{tabular}
\end{table}

If the impedance is the goal of the validation, the numerically resulting surface impedance is compared with the analytical solution according to the Miki model \cite{Miki1990,RichardFernandezGrandeBrunskogEtAl2017}.
Therein, the characteristic impedance $Z_c$ is given by
\begin{eqnarray}
    Z_c &=& \varrho c \left(1 + 0.070 \left(\frac{f}{\sigma}\right)^{-0.632} \right. \\
    &-& \left.0.107\sqrt{-1} \left(\frac{f}{\sigma}\right)^{-0.632}\right) \nonumber
\end{eqnarray}
with $c$ as the speed of sound, $f$ as frequency, and $\sigma$ as flow resistivity.

Assuming a plane wave the surface impedance of porous material backed by another material is given by \cite{PindJeongEngsigKarupEtAl2020} as
\begin{eqnarray}
    Z_s(\phi) = \dfrac{Z_c k_t}{k_x}\left( 
    \dfrac
    {-\sqrt{-1}Z_b\cot (k_x l) + Z_c \frac{k_t}{k_x}}
    {Z_b - \sqrt{-1} Z_c \frac{k_t}{k_x} \cot (k_x l)}
    \right).
\end{eqnarray}
Therein, $l$ denotes the thickness of the porous material, and $\phi$ is the incidence angle with respect to the porous surface. 
The wavenumber $k_t$ is defined as
\begin{eqnarray}
    k_t(f) &= \dfrac{2\pi f}{c} \left( 1 + 0.109\left( \dfrac{f}{\sigma} \right)^{-0.618} \right. \\ \nonumber & \left. -0.160\sqrt{-1}\left( \dfrac{f}{\sigma}\right)^{-0.618}  \right)
\end{eqnarray}
with $k_x = \sqrt{k_t^2 - k^2\sin^2(\phi)}$ and $k$ as free space wavenumber.

In case the porous material is backed by a rigid wall, see Fig.~\ref{fig_a1_a2_setup} (left), the surface impedance reduces to
\begin{equation}
    Z_s(\phi) = -\sqrt{-1}Z_c\left(\dfrac{k_t}{k_x}\right)\cot(k_xl).
    \label{eq_z_s_backed_by_rigid_wall}
\end{equation}
If there is an air cavity between the porous material and the wall, see Fig.~\ref{fig_a1_a2_setup} (center), 
\begin{equation}
    Z_b = -\varrho c \sqrt{-1} \dfrac{\cot(k_0 \cos(\phi)l_0)}{\cos(\phi)} 
\end{equation} 
holds with $l_o$ as the thickness of the cavity.

\subsection{\label{ssec_1d_absorber} 1D - absorber}

As first validation cases, the absorber configurations A1 and A2 are examined. 
The 1-D computational domain with a length of 2.5 m is discretized with 626 equidistantly distributed points. 
An explicit 4th order derivative is used. 
An explicit 4th order Runge-Kutta method is used for temporal integration. 
20990 time steps are calculated at a sampling rate of 96 kHz. 
The resulting CFL number is 0.89. 
Characteristic non-reflecting boundary conditions \cite{Thompson1987,PoinsotLele1992} are imposed on both sides of the computational domain.
An acoustic source is imposed at the coordinate $x_1 = 0.4$ m during the entire computational time. 
Its signal is given by a linear chirp from 50 to 3500 Hz over the simulation time. 

Both absorber configurations are modeled by a suitable choice of the effective volume $\phi$ and the Darcy penalization $\chi$. 
The functions are given by:
\begin{align}
\phi(x) = 1  
     &- \dfrac{(1-\phi_p)}{2}  \left[(\tanh((x-s_p) / \delta))\right.       \\
     &\qquad\qquad- \left.(\tanh((x-e_p) / \delta) )\right]                           \nonumber \\
     &- \dfrac{(1-\phi_w)}{2}  \left[(\tanh((x-s_w)/\delta) ) \right.     \nonumber \\
     &\qquad\qquad- \left.(\tanh((x-e_w)/\delta) )\right]                             \nonumber 
\end{align}
and
\begin{align}
    \chi(x) &= + \dfrac{a_p}{2}
             \left[(\tanh((x-s_p)/\delta)) \right.       \\
            &\qquad\quad- (\tanh((x-e_p  )/\delta)))                 \nonumber\\
            &\qquad\quad+  ((\tanh((x-s_w)/\delta))                  \nonumber\\
            &\qquad\quad- \left.(\tanh((x-e_w  )/\delta))  \right].   \nonumber
\end{align}
The values for the spatial parameters start $(s)$ and end $(e)$ of the absorber material $\bullet_p$ and the backing wall $\bullet_w$ are stated in Tbl.~\ref{tab_a1_a2_penalization_values}.
The steepness of the $\tanh$-flanks is controlled using a value of $1.5 \Delta x$ for the parameter $\delta$.
Please note that in general, $\delta$ can be chosen independently for both functions $\phi$ and $\chi$. 
 
\begin{table}[tb]
\caption{
\label{tab_a1_a2_penalization_values}
Parameters of effective volume and Darcy penalization for configurations A1 and A2.\bigskip}
\footnotesize
\centering
\begin{tabular}{cccc|cccc}
\hline
\hline
~ & \multicolumn{3}{c}{\tiny{effective volume}} & \multicolumn{4}{c}{\tiny{Darcy penalization}} \\
~ & $s_w$ & $e_w$ & $\phi_w$ & $s_p$ & $e_p$ & $a_p$ & $\phi_p$  \\
\hline
A1 & $2.115$ & $\infty$ & $10^{-6}$ & $2.0$ & $2.100$  &  $4300$ & $1.0$\\
A2 & $2.212$ & $\infty$ & $10^{-6}$ & $2.0$ & $2.056$  & $12500$ & $1.0$\\
\hline
\hline
\end{tabular}
\end{table}

\begin{figure}
    \centering
    \includegraphics[width = \linewidth]{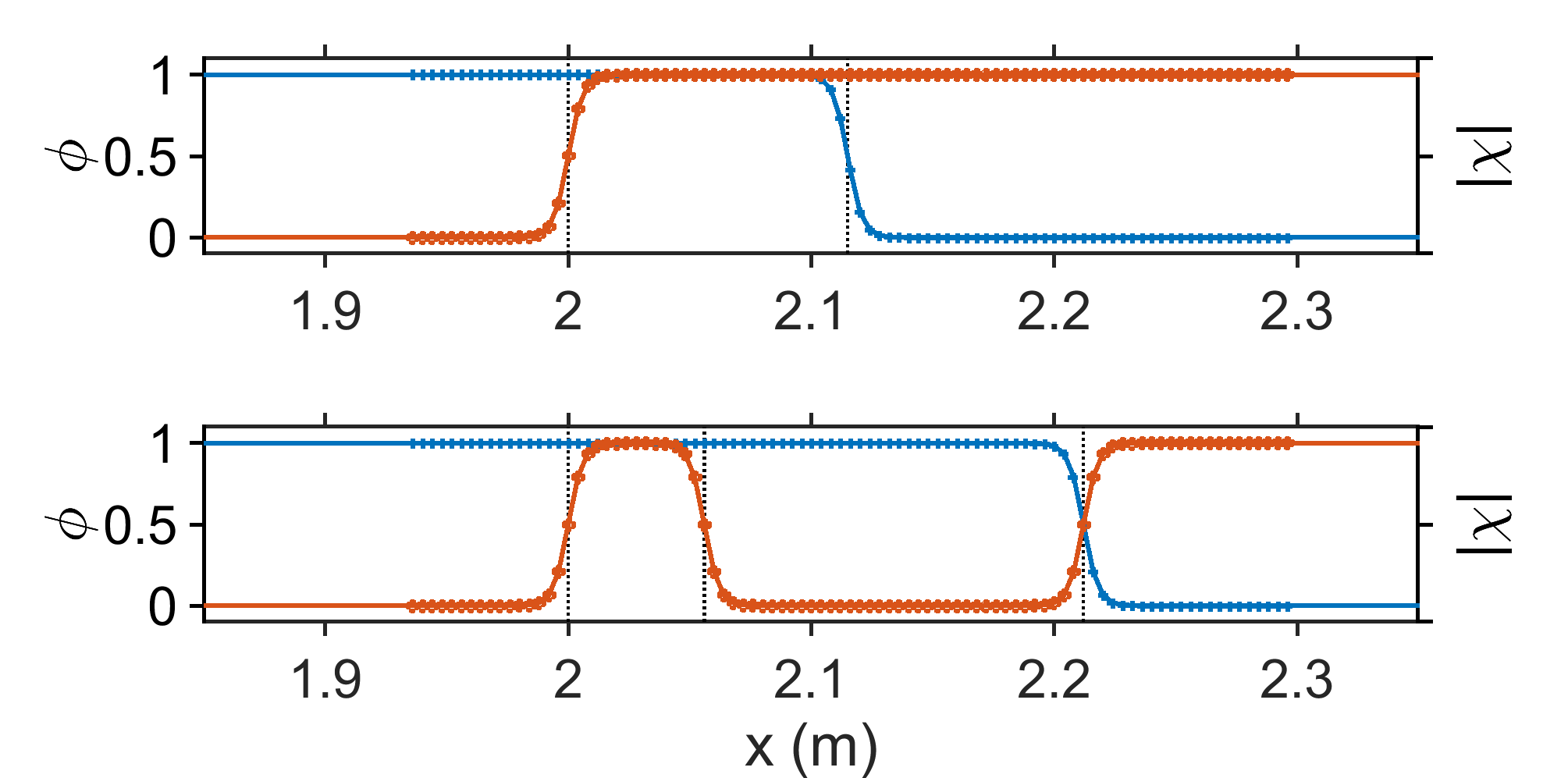}
    \caption{(color online) Resulting curves for the effective volume ($\phi$, blue) and the penalization ($\chi$, red) for the cases A1 (top) and A2 (bottom).
    The course for $\chi$ is normalized with the respective amplitude $a_p$.
    The vertical lines correspond to turning points in the profiles.
    The markers correspond to the grid points in the finite-difference grid used.
    }
    \label{fig_zs_absorber_setup}
\end{figure}

The resulting curves for the effective volume $\phi$ and the Darcy term are shown in Fig.~\ref{fig_zs_absorber_setup}.
 
The values are adjusted manually to achieve a good match with the validation reference.
A detailed optimization was not carried out.
The values are similar to the reference values.

\begin{figure}
    \centering
    \includegraphics[width = \linewidth]{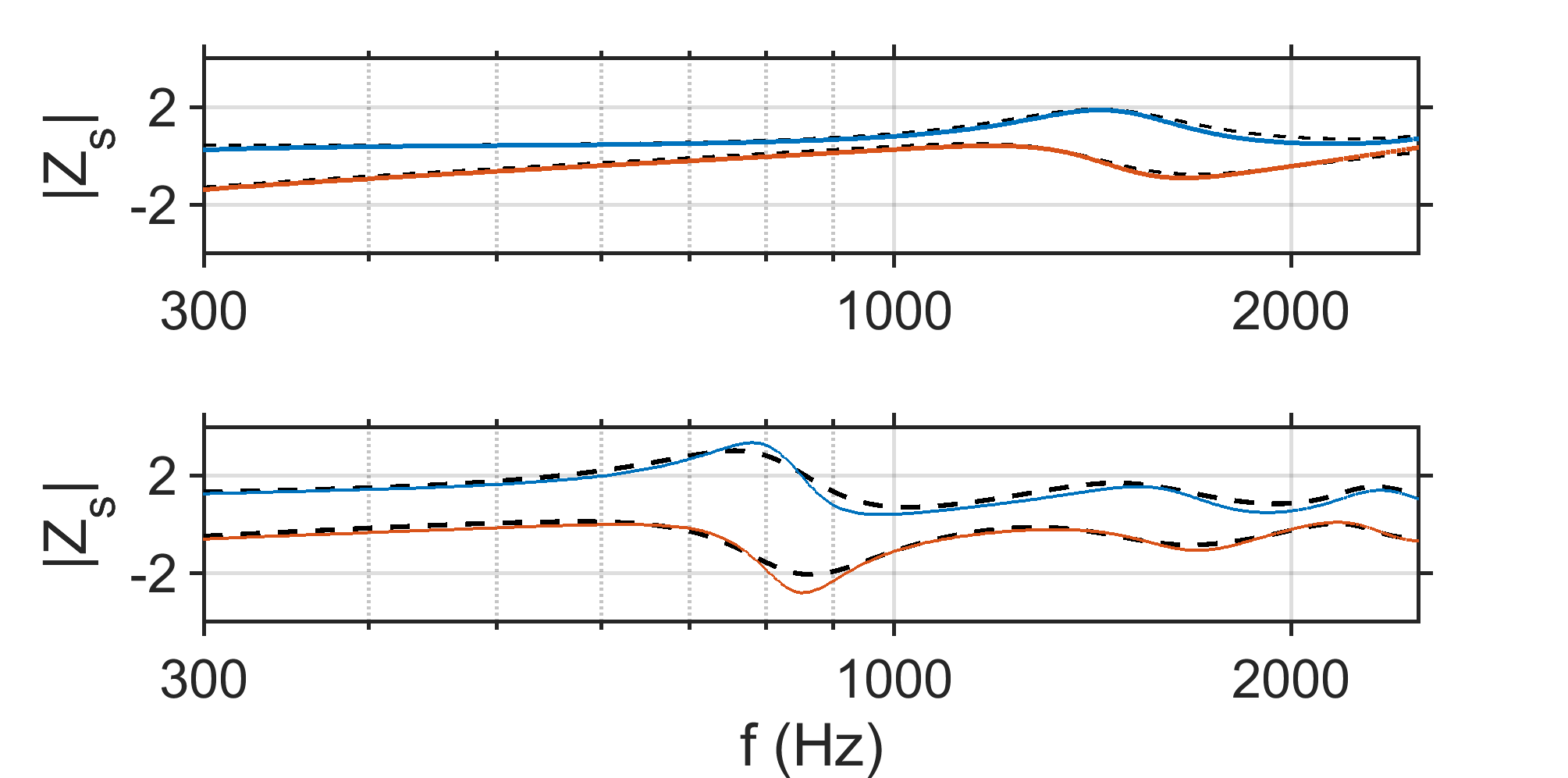}
    \caption{(color online) Resulting surface impedance $Z_s$ normalized with $\varrho c$ for the cases A1 (top) and A2 (bottom) - real (blue) and imaginary part (red) compared to the analytical solution according to the Miki model.}
    \label{fig_zs_absorber}
\end{figure}

The resulting (normal incidence) surface impedance is computed using $Z_s = {\hat p}/{\hat u}$, where $\hat p$ and $\hat u$ are the Fourier transformed pressure and velocity time responses at position $x = s_p$.
In comparison to the impedance based on the Miki-model a good agreement is found, see Fig.~\ref{fig_zs_absorber}.
 
We conclude that the Brinkman penalization approach can mimic acoustic absorber configurations very well with a small adaption of the modeling parameters. 

\subsection{\label{ssec_1d_helmholtz} 1D - responsive surface}

The numerical case B considers a responsive surface.
The underlying idea is that the effective volume of the porosity can be interpreted as a channel with varying cross-sections \cite{AndrianovWarnecke2004,Reiss2021}. 

The 1-D computational domain with a length of 2.5 m is discretized with 1252 equidistantly distributed points. 
Again, an explicit 4th order derivation scheme and an explicit 4th order Runge Kutta method are used for discretization. 
20990 time steps are calculated at a sampling rate of 192 kHz. 
The resulting CFL number is 0.89. 
Characteristic non-reflecting boundary conditions are imposed on both sides of the computational domain.
An acoustic source is imposed at the coordinate $x_1 = 0.2$ m during the entire computational time. 
As before, its signal is given by a linear chirp from 50 to 3500 Hz over the computational time. 

\begin{figure}
    \centering
    \includegraphics[width = \linewidth]{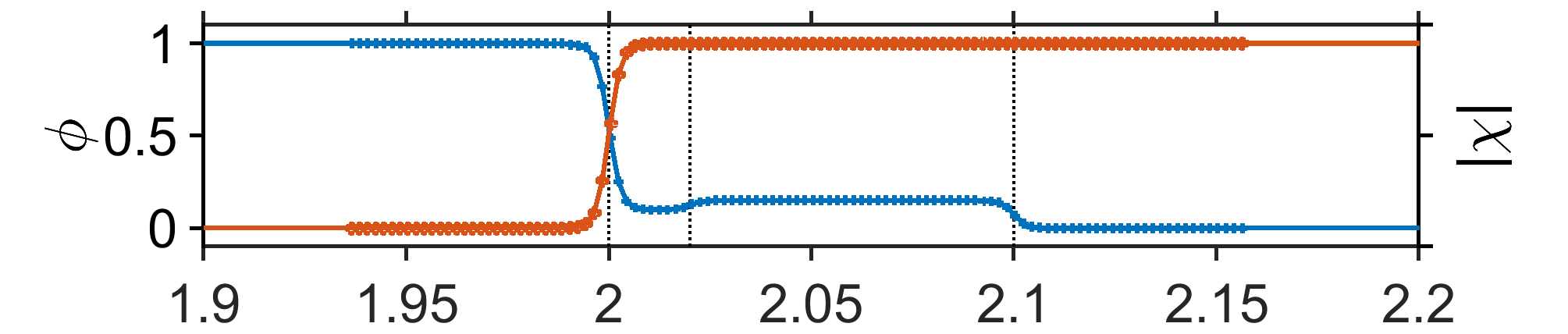}
    \caption{(color online) Resulting curves for the effective volume ($\phi$, blue) and the penalization ($\chi$, red) for the reactive surface B.
    The course for $\chi$ is normalized with respect to the maximal amplitude.
    The vertical lines correspond to turning points in the $\tanh$-functions, defining the neck and the resonator volume.
    The markers correspond to the grid points in the finite-difference grid used.}
    \label{fig_helmholtz_setup}
\end{figure}

The effective volume and the Darcy penalization are chosen to mimic an acoustic element containing a Helmholtz resonator.
In detail, $\tanh$-functions are used to model the neck and the resonator volume of the acoustic element.
A value of $\delta = 1.5\Delta x$ is again chosen as the steepness of the flanks of $\phi$ and $\chi$.
The neck starts at 2 m and ends at 2.02 m, where the resonator volume with a length of 0.08 m volume starts.
Behind the resonator, the wall is modeled to begin at 2.1 m.
Within the neck, the effective volume is set to 0.1, while it is increased for the resonator volume to 0.15.
To mimic the wall, $\phi$ is set to $10^{-6}$ in the corresponding wall region.
The Darcy penalization is chosen to a maximum value of 250 starting at the neck.
The resulting curves for the effective volume and the penalization are shown in Fig.~\ref{fig_helmholtz_setup}.
 
The resulting acoustic properties of the model are validated by comparison of the impedance 
$Z_s = {\hat p}/{\hat u}$ at the neck entry and the analytic impedance of an Helmholtz resonator given by \cite{Ehrenfried2004}
\begin{equation}
    Z_{s,HR} = R_l + \sqrt{-1}\dfrac{\varrho H}{\omega S}\left(\omega^2 - \dfrac{c^2S}{VH}\right)
\end{equation}
or the corresponding reflection coefficient $R = (Z/(\varrho c) - 1)/(Z/(\varrho c) + 1)$ respectively.
Therein, $R_l$ denotes an additional damping coefficient, $H$ the neck length, $S$ the diameter of the neck tube, and $V$ the volume of the resonator.
The corresponding values for the case examined, here, are shown in Tbl.~\ref{tab_r1_values}.

\begin{table}[tb]
\caption{
\label{tab_r1_values}
Parameters of the model for an acoustic element used to describe configuration B.\bigskip}
\centering
\begin{tabular}{cccc}
\hline
\hline
$H$ & $S$ & $V$  & $R_l$  \\
\hline
0.0367 &  0.0205 &     0.0025 &        1850 \\
\hline
\hline
\end{tabular}
\end{table}

As before, the parameters have been adjusted to obtain a good match.
A detailed optimization was not carried out.

\begin{figure}
    \centering
    \includegraphics[width = \linewidth]{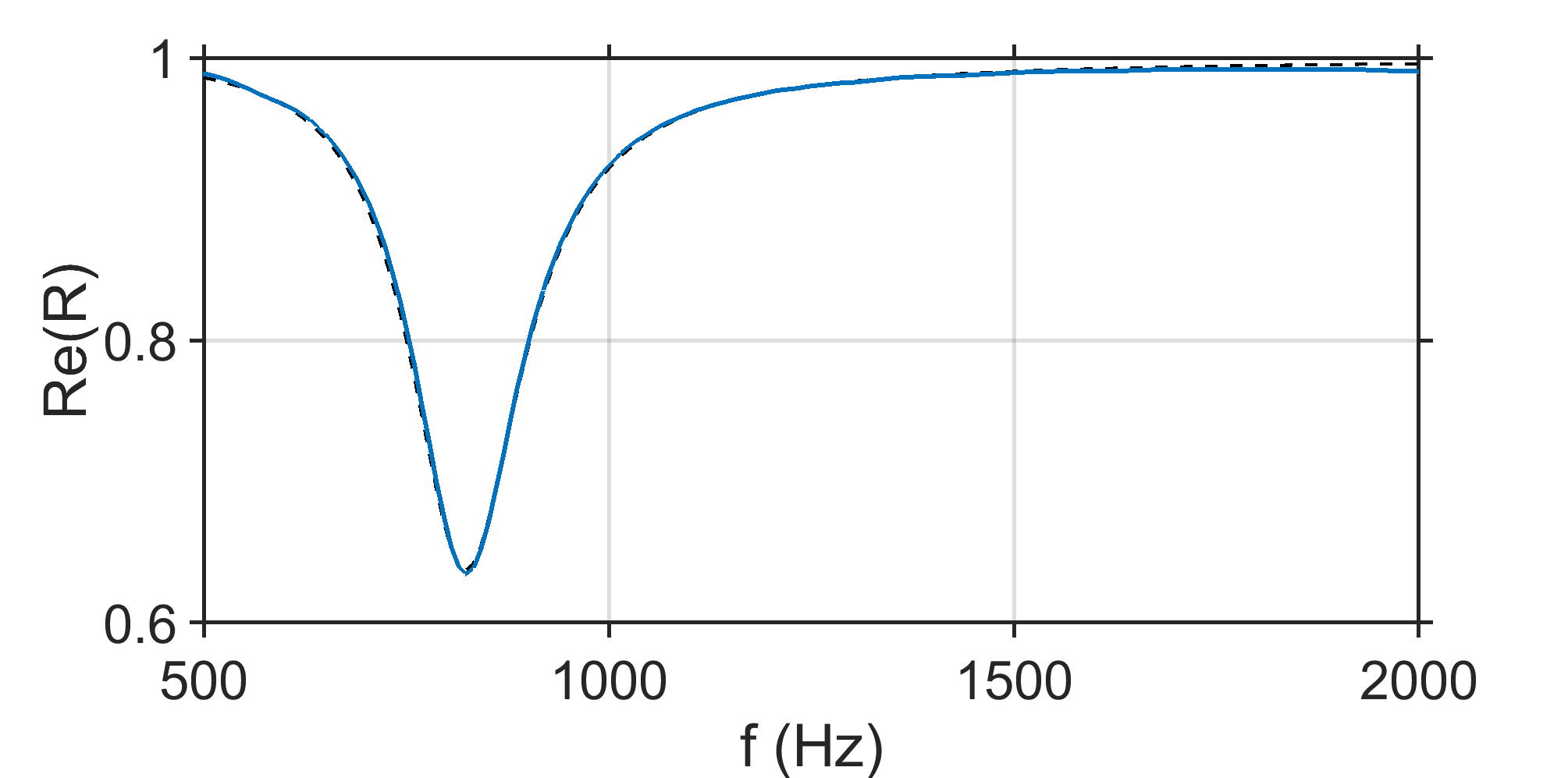}
    \caption{(color online) Progress of the reflection coefficient (blue) for configuration B compared to the analytical solution of an acoustic element (black, dashed).}
    \label{fig_r_helmholtz}
\end{figure}

Please note that the values in the 1-D-model examined here have no direct physical equivalence.
Nevertheless, the comparison of the Brinkman penalization approach and the analytical solution shows a very good agreement, see Fig.~\ref{fig_r_helmholtz}.
 
The Brinkman penalization can mimic responsive boundary conditions.
However, in the course of the manuscript, only passive wall impedance boundary conditions will be examined further.

\subsection{\label{ssec_2d_circ} 2D - Circular Domain}

In the following, a 2-D circular domain is considered.
The dimensions of the computational domain are $0 \leq  x_1 \leq 1.4$ m and $0 \leq x_2 \leq 1.4$ m.
A uniform grid with $145 \times 145$ points is used.
A fourth-order accurate symmetric derivation stencil is employed. 
The computational time of 3 s is separated into $144000$ time steps which result in a CFL condition of 0.735.
An explicit fourth-order Runge-Kutta scheme is used for the time integration.
All domain-boundaries are treated as non-reflecting using characteristic boundary conditions \cite{Thompson1987, PoinsotLele1992}.
Since the boundary condition is in the modeled wall no substantial influence on the results is expected. 
To ensure stability, an implicit filter of 4th order is employed at each time step \cite{GaitondeVisbal2000}.
The simulation is initialized with a Gaussian pulse with a spatial variance of 0.05 at $x_1 = 1.0$ m and $x_2 = 0.8$ m.

\begin{figure}
    \centering
    \includegraphics[width = \linewidth]{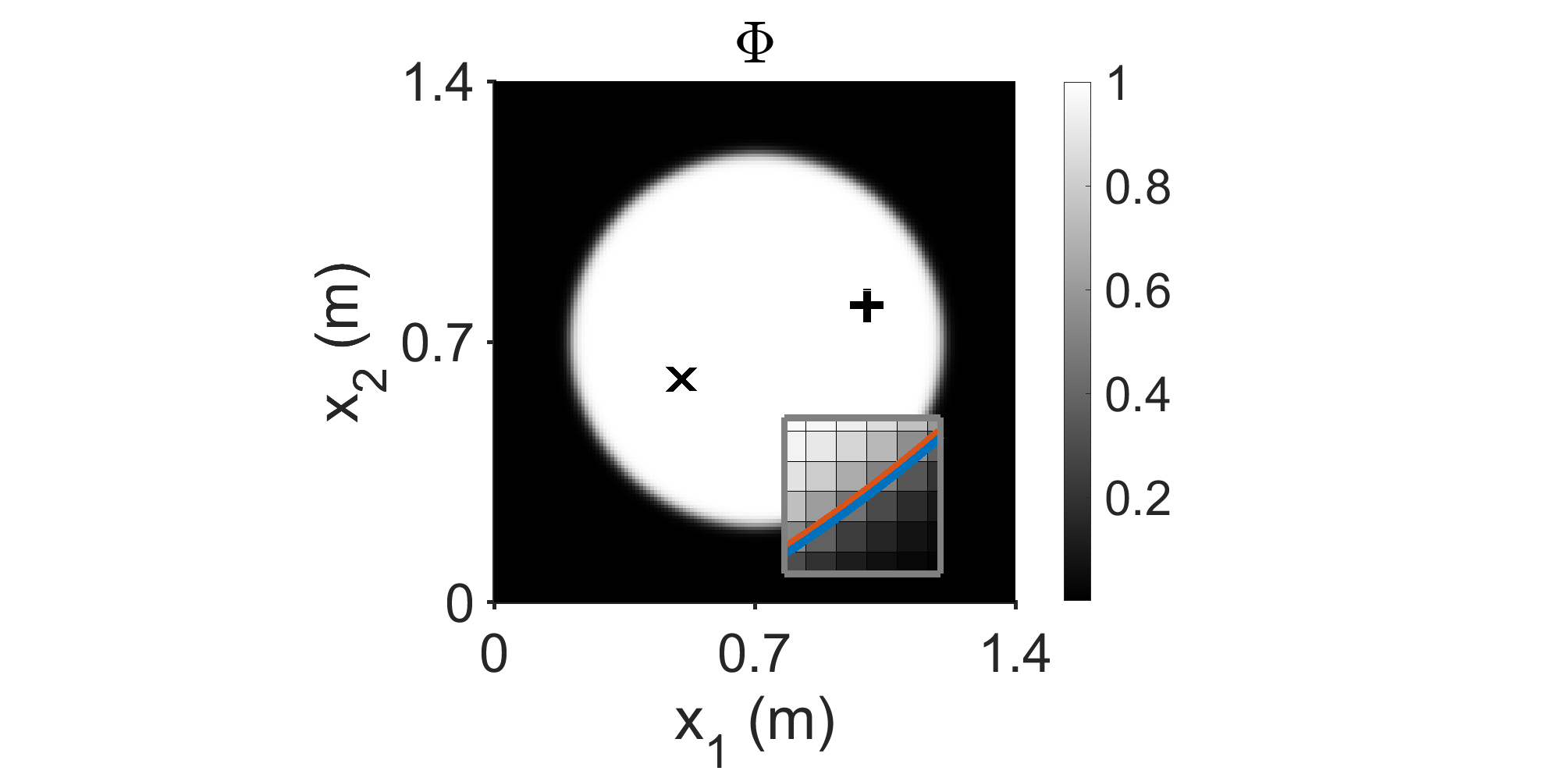}
    \caption{(color online) Spatial distribution of the effective volume $\phi$ in setup C.
    The source position and the receiver positions are marked by $+$ and $x$, respectively.
    The inset shows the $\phi$ distribution zoomed in the wall area.
    Turning point of the function selected at a distance of half a meter from the center (red).
    Resulting room size for a good match of the room modes (blue), see text for more details.
    }
    \label{fig_ring_setup}
\end{figure}

The circular geometry is modeled by the Brinkman penalization approach, see Fig.~\ref{fig_ring_setup}.
The effective volume $\phi$ is varied between $\phi_p$ and 1 using the following formula
\begin{equation} 
    \phi(r) = 1 - \dfrac{(1-\phi_p)}{2}  \left(\tanh\left(\dfrac{r - 0.5}{\delta}\right) + 1\right).
\end{equation}
Therein $\phi_p = 10^{-5}$ denotes the minimal value of the effective volume and $r$ the radial distance of each grid point to the center of the domain.
A value of $\delta = 1.75\Delta x$ is chosen as the steepness of the flank.
There is no Darcy penalization so that $\chi = 0$ applies to the entire domain.
 
\begin{figure}
    \centering
    \includegraphics[width = \linewidth]{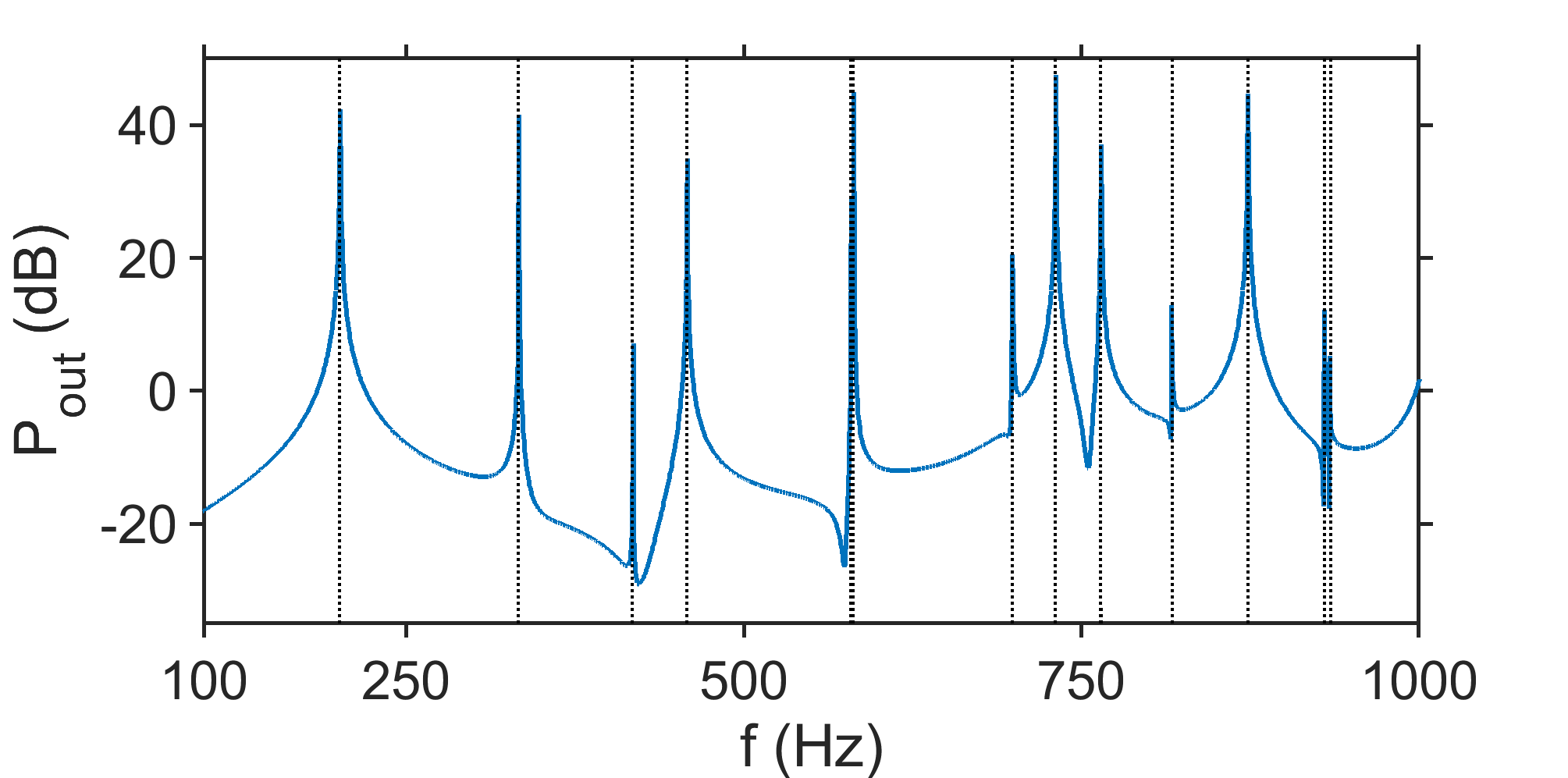}
    \caption{(color online) Frequency responses (blue) in the 2-D-circular C domain using the volume penalization approach. Analytic modes are presented by vertical lines (black, dashed), calculated using Green’s function. 
    Note that also double peaks are slightly visible.
    }
    \label{fig_ring_modes}
\end{figure}

A comparison of the resulting frequency responses, evaluated at the spatial location $x_1 = 0.5$ m and $x_2 = 0.6$ m with the analytical ones, given by roots of the derivative of the Bessel function \cite[p.~110f]{JacobsenJuhl2013}, shows a good agreement, see Fig.~\ref{fig_ring_modes}.
Also, frequencies that are very close together are correctly identified and separated.
However, the analytical solution must be corrected to obtain the results.
The analytical radius is increased by 0.2\% of the actual dimension.
The deviation is due to the approximate character of the penalization approach.
However, the deviation is small and corresponds to about $0.1 \Delta x$.
 
The Brinkman penalization can model non-grid-aligned geometries with good quality and, in particular, without complex adaptations of the computational grid as in \cite{PindEngsigKarupJeongEtAl2019}.
The geometry is fully encoded in the values of $\phi$ and $\chi$ which are created once at simulation start. 
Thus, complex rooms are modeled without the need to evaluate surface elements in every time step which can be expensive for complex structures \cite{Vorlaender2013}.

\subsection{\label{ssec_3d_room 3D} 3D - reverberation chamber}

For this case, a three-dimensional cubic domain is considered.
The dimensions of the computational domain are $-0.5 \leq  x_i \leq 1.5$ m for all three spatial directions.
A uniform grid with $201 \times 201 \times 201$ points is used.
For the spatial discretization, a fourth-order accurate implicit symmetric derivation stencil is used \cite{Lele1992}. 
The computational time of 3 s is separated into $144000$ time steps which result in a CFL condition of 0.715.
An explicit fourth-order Runge-Kutta scheme is employed for the time integration.
All domain-boundaries are treated as non-reflecting using characteristic boundary conditions \cite{Thompson1987, PoinsotLele1992}.
The simulation is initialized with a Gaussian pulse with a full width at half maximum of $2 \Delta x_i$ at $x_1 = 0.25$ m, $x_2 = 0.75$ m and $x_3 = 0.60$ m.

\begin{figure}
    \centering
    \includegraphics[width = \linewidth]{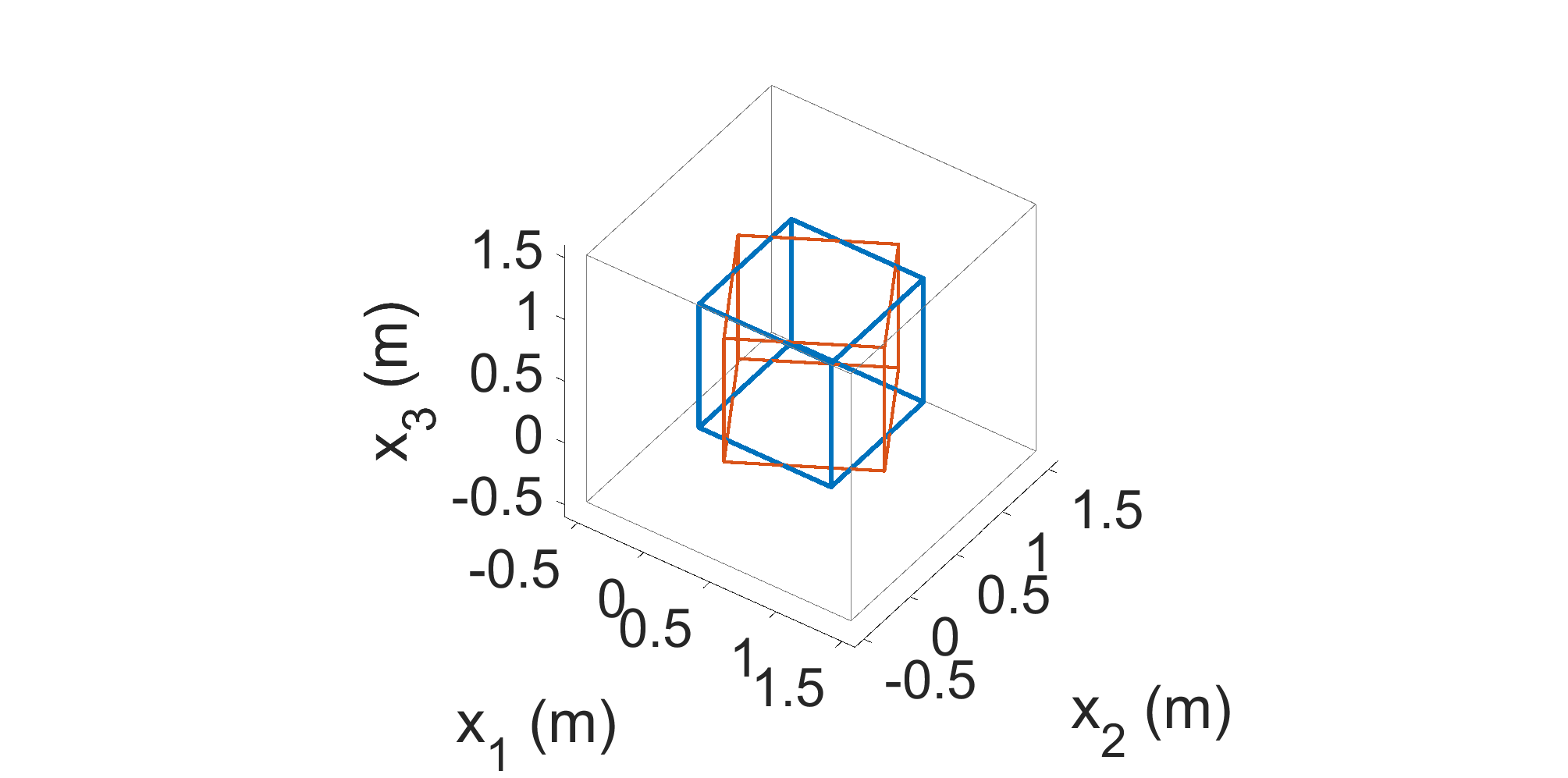}
    \caption{(color online) Edges of the room D modeled by the effective volume approach.
    Edges parallel to the computational grid (blue).
    Edges rotated by $30^\circ$ with respect to the z-axis with the center point unchanged (red).}
    \label{fig_room_setup_rotation}
\end{figure}

The room geometry is modeled using the effective volume $\phi$ only.
No Darcy penalization is applied.
Thus, $\chi=0$ applies in the whole computational domain.
The function $\phi$ is varied between a minimum value of $10^{-5}$ and 1 employing $\tanh$-functions as above.
The functions define a rectangular room with the dimensions $1 \times 1 \times 1$ meters, centered in the computational domain.
The locations of the boundary lines are shown in Fig.~\ref{fig_room_setup_rotation}. 
A value of $\delta = 1.75\Delta x$ is chosen as the steepness of the flanks.
Two cases are investigated. 
In the first case, the edges of the room and the computational grid are parallel to each other.
In the second case, the room is rotated by 30$^\circ$ with respect to the $x_3$-axis, whereby the center of the room remains unchanged.

\begin{figure}
    \centering
    \includegraphics[width = \linewidth]{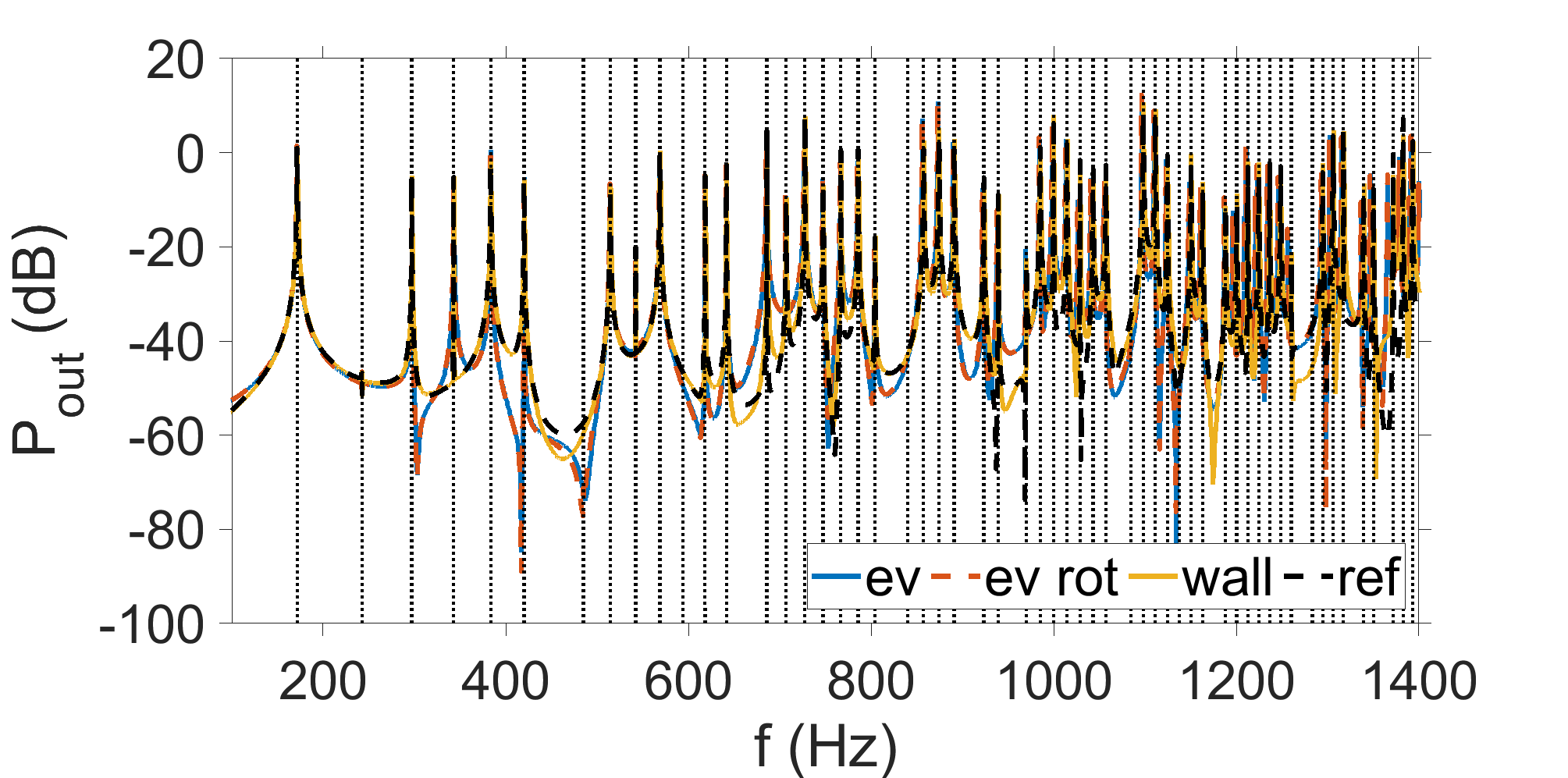}
    \caption{(color online) Frequency responses in the square 3-D domain (D) under consideration.
        FDFT-domain results using rigid boundary conditions (yellow),
        using the effective volume (blue), the effective volume under $30^\circ$ rotation (red, dashed), and the analytical solution (black, dashed).
        The analytical eigenmodes are shown by vertical lines (black, dotted).
        }
    \label{fig_room_spectra}
\end{figure}

A comparison of the resulting frequency responses, evaluated at the spatial location $x_1 = 0.85$, $x_2 = 0.30$ and $x_3 = 0.80$ m with the analytical ones \cite{Sakamoto2007} and a simulation using fully reflecting boundary conditions show a very good agreement, see Fig.~\ref{fig_room_spectra}.
That applies in particular to the rotated room, which shows that no alignment of the grid to the immersed boundary is necessary. 
Again, the room size in the analytical solution must be adapted to obtain the present result.
The edge lengths of the analytical room are assumed to be 0.2\% longer.
The deviation corresponds to about $0.2 \Delta x_i$. 
A similar offset was observed in \cite{Reiss2021} and could be compensated in the setup of the simulation.
 
The Brinkman penalization approach enables the simulation of three-dimensional acoustic configurations.
An alignment of the penalization concerning the computational grid is not necessary.

\subsection{\label{ssec_3d_angle} 3D - absorber - angle dependence}
In the following, the angle dependency of the surface impedance of configuration A1 using the Brinkman penalization is examined.
A 3-D domain is considered.
The dimensions of the computational domain are $-0.5 \leq  x_1 \leq 6.5$ m, $0.0 \leq  x_2 \leq 10.0$ m and $0.0 \leq  x_3 \leq 10.0$ m.
A uniform grid with $700 \times 1000 \times 1000$ points is used.
For the spatial discretization, a fourth-order accurate implicit symmetric derivation stencil was used \cite{Lele1992}. 
The computational time is separated into $1500$ time steps using a step width of $\Delta t = 1/48000$ s.
The CFL condition is 0.715.
An explicit fourth-order Runge-Kutta scheme is employed for the time integration.
All domain-boundaries are treated as non-reflecting using characteristic boundary conditions \cite{Thompson1987, PoinsotLele1992}.
The calculation is carried out in parallel on 80 cores, using the computational aeroacoustic framework presented in \cite{LemkeStein2021}. 
The simulation is initialized with a pressure Gaussian pulse with a sigma of $0.05$ at $x_1 = 3.5$, $x_2 = 3.5$ and $x_3 = 5.0$ m.

\begin{figure}
    \centering
    \includegraphics[width = \linewidth]{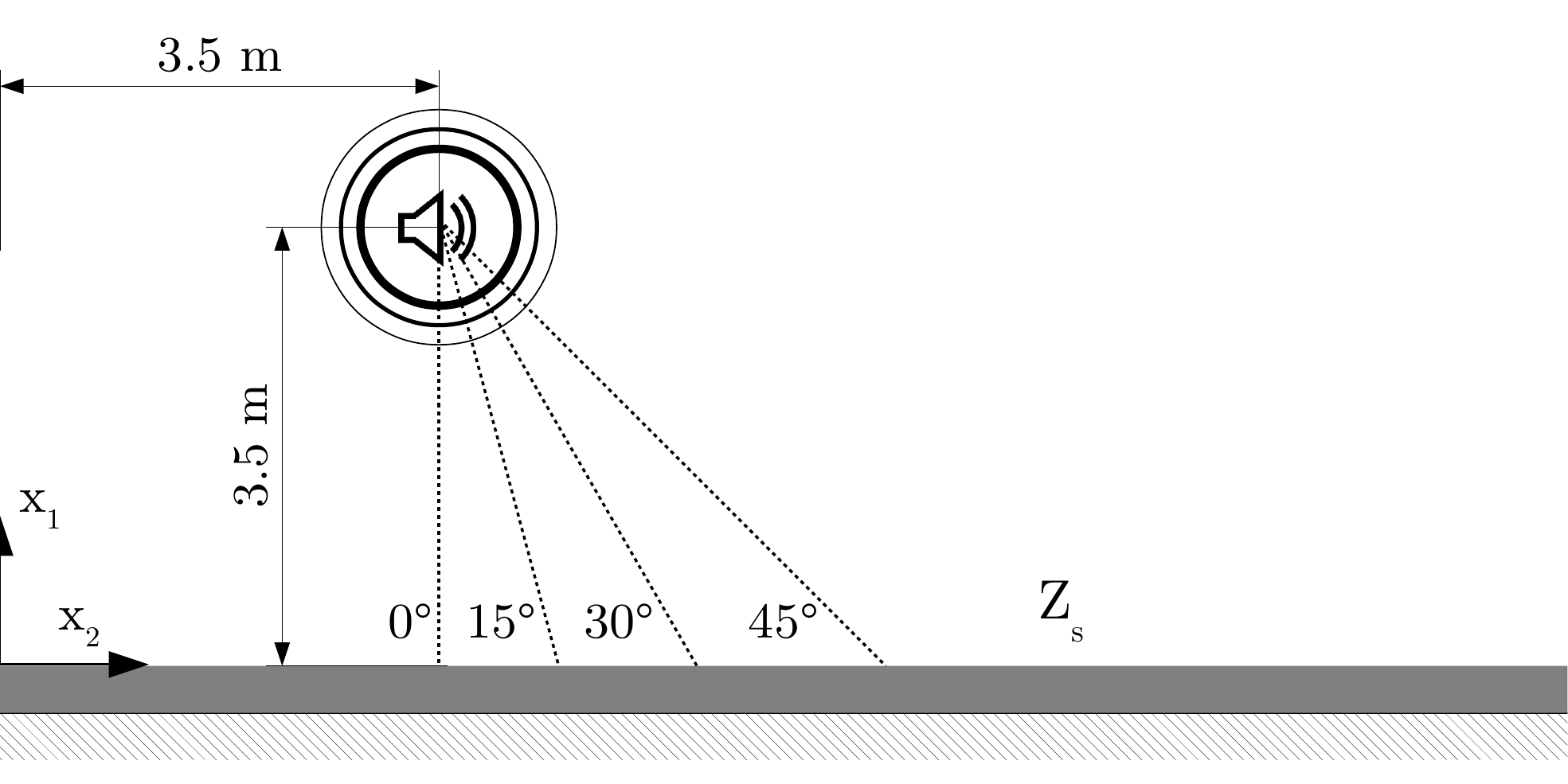}
    \caption{2-D-section of the three-dimensional computational domain E at $x_3 = 5$ m.}
    \label{fig_principle_absorber_angle}
\end{figure}

The absorber is modeled using the effective volume $\phi$ and the Darcy penalization $\chi$.
The courses of the function correspond to the case A1, shown in Fig.~\ref{fig_zs_absorber_setup} (top).
The surface of the porous material is located at $x_1 = 0$ m, see Fig.~\ref{fig_principle_absorber_angle}.

\begin{figure}
    \centering
    \includegraphics[width = \linewidth]{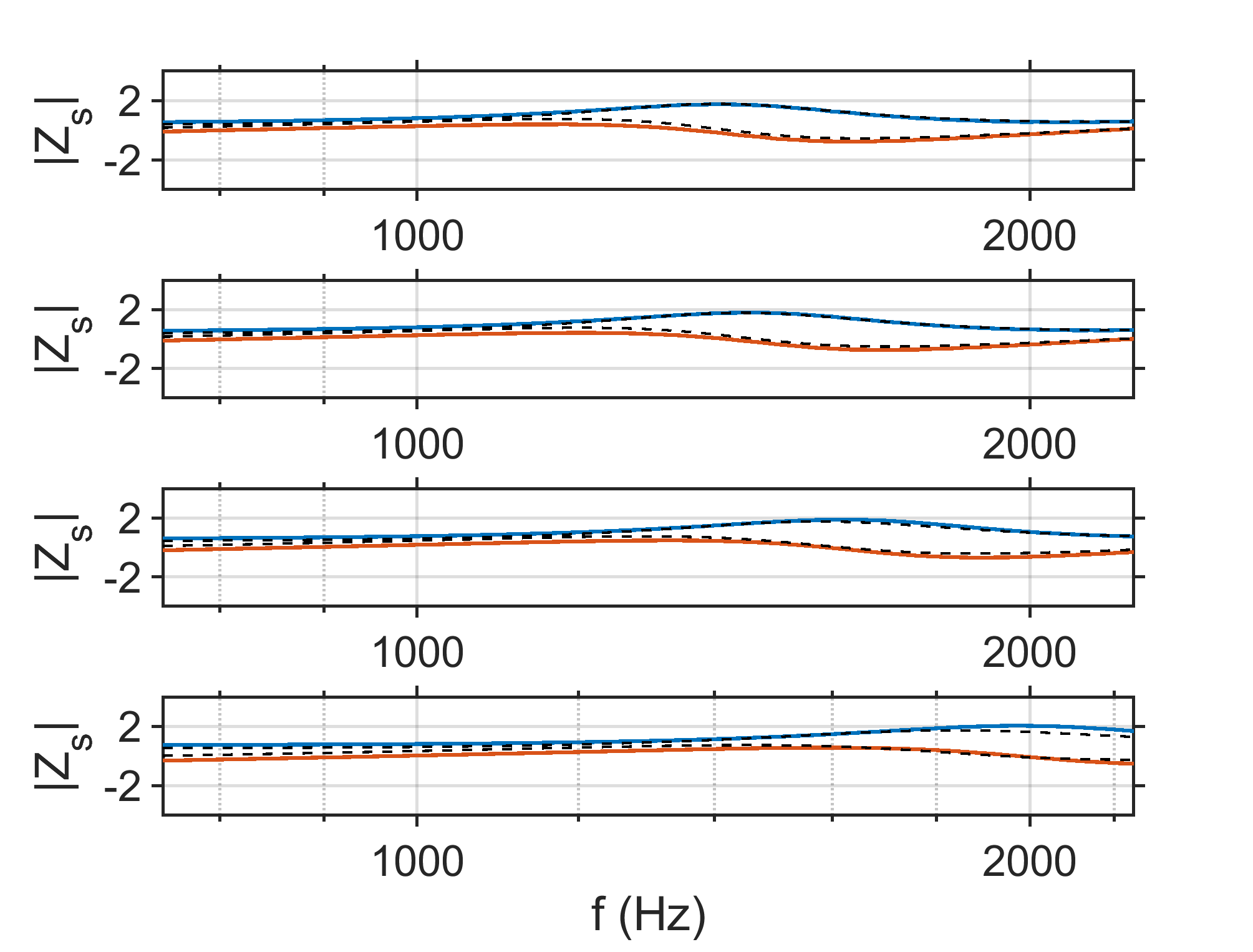}
    \caption{(color online) Resulting surface impedance $Z_s$ for different incidence angles in setup E (top to bottom: 0$^\circ$, 15$^\circ$, 30$^\circ$, 45$^\circ$) normalized with $\varrho c$ - real (blue) and imaginary part (red) compared to the analytical solution (black, dashed) according to the Miki model.}
    \label{fig_zs_angle_45}
\end{figure}

To analyze the angle dependence of the penalization method, the wall impedance is evaluated at different points on the absorber surface at $x_1 = 0$,  see Fig.~\ref{fig_principle_absorber_angle}.
The impedance $Z_s = {\hat p}/{\hat u}$ is compared with the analytical model \eqref{eq_z_s_backed_by_rigid_wall}.
Corresponding curves for four angles from $0$ to $45^\circ$ are shown in Fig.~\ref{fig_zs_angle_45}.
A good match is found.
Please note that plane waves are assumed in the reference model used, which is only approximated in the performed numerical simulation.

The Brinkman penalization can mimic the angle dependence of the wall impedance of a typical absorber configuration.

\section{\label{sec_summary} Summary}

A Brinkman penalization method for the approximate representation of acoustic wall boundary conditions was presented.
The approach allows the realization of damping as well as responsive boundary conditions.
The effective volume and the Darcy penalization are easy to interpret.
The approach is fully parallelizable and easy to implement in existing time-domain simulation codes.
Compared to other methods, there is no need for complex grid adaptations.
The use of the effective volume does not lead to any significant limitation of the time step. 
The same applies to the Darcy terms as long as typical acoustic configurations are considered.
The location of boundaries is changed by the smoothing of the modeling functions.
However, a simple adjustment in the context of 1-D examinations is possible and can be transferred to 2 and 3 dimensions.

Through the spatial distribution of the penalization terms, a large number of degrees of freedom is available which allows the model to be adapted to experimental data as well as the simple optimization of impedance wall boundary conditions.
The latter is the goal of future research.

\section*{Acknowledgments}
The authors acknowledge financial support by the Deutsche Forschungsgemeinschaft (DFG) within the project LE 3888/2.

\appendix
\section{Acoustic Equations}

The acoustic equations are derived from the Euler equations (\ref{eqn_mass}-\ref{eqn_momentum}, \ref{eqn_pressure}) by linearization. 
Inserting $ p= p_0 +p'$,  $u= u_0+u'$ and $ \rho = \rho_0 + \rho'$, and keeping only terms linear in the fluctuations, yields
\begin{align}
\bm{\phi}	\rho_0 \partial_t (  u')  + & \phantom{p_0 } \bm{\phi} \partial_x ( p')  &=&  \bm{\phi}\bm{ \chi (u^t - u')}  \label{acoustic_mom} \\  
 	\bm{\phi} \partial_t ( p' ) +&  p_0 \gamma \partial_x( \bm{\phi}  u'   ) 
 	              	                &=& 0,  \label{acoustic_p} 
\end{align}
where $u_0=0 $ and constant $\rho_0$ and $p_0 $ are assumed.  
The mass equation \eqref{eqn_mass} can be ignored if adiabatic flows are assumed, as usually done for acoustic applications. 
The two equations can be combine with $\partial_t \phi =0 $ to 
\begin{align}
 	\bm{\phi} \partial_t^2 ( p' ) - c^2   \partial_x ( \phi  \partial_x(  p'   ) ) 
 	              	                &= - \partial_x \bm{\phi}\bm{ \chi (u^t - u')}    
\end{align}
This reveals that the usual wave equation with the wave velocity $c$ is recovered in areas with constant $\phi$. 
Rewriting the term 
$
\partial_x ( \phi  \partial_x(  p'   ) ) 
= \partial_x (  \partial_x(  p'   ) ) 
+(  \partial_x  p'   ) (\partial_x  \phi  ) 
$ 
allows to obtain a standard Laplacian. 
For computational purposes the form (\ref{acoustic_mom},\ref{acoustic_p}) is preferred as it avoids second derivatives.

\bibliographystyle{abbrv}
\bibliography{bib}

\end{document}